\documentclass[aps,prl,twocolumn,groupedaddress]{revtex4}
\usepackage{graphicx}
\usepackage{epsfig}
\usepackage{float}

\newcommand{\be}{\begin{equation}}
\newcommand{\ee}{\end{equation}}
\begin{document}

\title{Nesting, spin-fluctuations, and odd-gap superconductivity in 
Na$_x$CoO$_2\cdot y $H$_2$O} 
\author{M.D. Johannes, I.I. Mazin, D.J. Singh, and D.A. Papaconstantopoulos}
\affiliation{Code 6391, Naval Research Laboratory, Washington, D.C. 20375}

\begin{abstract} We have calculated the one-electron susceptibility of hydrated Na$_x$CoO$_2$ and find strong
nesting nearly commensurate with a 2$\times$2 superstructure. The nesting involves about 70\% of all electrons
at the Fermi level and is robust with respect to doping. This nesting creates a tendency to a charge density
wave compatible with the charge order often seen at $x\approx 0.5$, which is usually ascribed to electrostatic
repulsion of Na ions. In the spin channel, it gives rise to strong spin-fluctuations, which should be important
for superconductivity.  The superconducting state most compatible with this nesting structure is an odd-gap
triplet s-wave state. \end{abstract}

\maketitle

{\it Introduction.} The recent discovery of superconductivity in the layered oxide Na$_{1/3}$CoO$_{2} \cdot
$1.4H$_{2}$O \cite{KTHS+03} is the subject of intense experimental and theoretical research, despite its
relatively low critical temperature, due to a combination of properties that suggest the possibility of a
non-trivial superconducting state and/or a non-trivial pairing mechanism. These include unusual magnetic,
thermodynamic, and transport properties, the apparent proximity to both structural and magnetic instabilities,
and the frustrated triangular Co lattice.  \cite{ITYS97,YWNSR+03,YANM+99,QHMLF+04,GB,SYL+04}

A variety of possible pairing interactions may be relevant in this system. Structural
instabilities have been reported in the non-hydrated compound \cite{BCS+04,QHMLF+04},
suggesting the possibility of a related soft mode and, correspondingly, strong
electron-phonon coupling. A high polarizability of water molecules may be responsible
for \textquotedblleft sandwich\textquotedblright\ type superconductivity, where the
paired electrons and the pairing bosons are spatially separated\cite{DAJB73}. Band
structure calculations yield a ferromagnetic ground state\cite{DJS00,DJS03} and 
favor an anti-ferromagnetic state over a non-magnetic one, presaging
long-range spin fluctuations that are observed at some dopings \cite{ATB+03},
Finally, and this is the central point of the current Letter, nesting
properties of the calculated Fermi surface suggest the existence of strong
anti-ferromagnetic spin fluctuations, which are bound to play an important role in
superconductivity (as well as in the normal transport). First and foremost, they make a
conventional singlet s-wave state rather unlikely.

Experimental evidence for a pairing symmetry is still
inconclusive. There are reports of a lack of reduction of the Knight shift
through the superconducting transition, suggesting a triplet state
with vector order parameter directed out of the plane \cite{WHKO+03, AKAK+03,
TWCM+}. Density of states probes: $\mu $SR,
NMR, and NQR have so far observed a non-exponential relaxation rate behavior below
$T_{c}$ \cite{TFGZ+,KIYI+}, inconsistent with a fully
gapped state; some reported a Hebel-Slichter maximum near
$T_{c}$\cite{YKMY,TWCM+}, suggesting a coherence peak in the DOS,
while others did not find such a maximum, possibly because of impurity
scattering.

Here, we calculate the one-electron susceptibility and show that it has strong structure in reciprocal space,
which is not related to crystal symmetry, but is, accidentally, nearly commensurate with the lattice. This
structure is robust with respect to doping and interlayer distance, and may even be responsible for the
reported superstructures (as opposed to an intuitive picture relating them solely to Coulomb ordering of Na
ions). The calculated spin fluctuations appear to be incompatible with either singlet or triplet BCS
superconductivity, but they are fully compatible with so-called odd-gap superconductivity, with the most
favorable symmetry being triplet $s$-wave. If this interaction is indeed responsible for the superconductivity,
then, somewhat similarly to MgB$_{2}$, superconductivity is driven by the Fermi surface pockets which have
relatively small volume (but large density of states).  It is the two-dimensionality of the electronic
structure that makes this possible.

{\it Electronic structure, nesting and susceptibility}.  The strength of Coulomb
correlations in Na$_{1/3}$CoO$_2$$\cdot$1.4H$_2$O (subsequently called NCO) is
yet unknown, but even in metals as strongly correlated as the high-T$_c$ cuprates, LDA
calculations consistently provide accurate Fermi surfaces.  The one-electron band
 structure of the parent compound, Na$_x$CoO$_2$, is by now well understood
\cite{DJS00,JKK-L03}.  The Co $d$-bands are split by the octahedral crystal field of the
surrounding oxygens into three $t_{2g}$ bands per layer, well separated from the two $e_g$
bands. The former further split in the hexagonal symmetry into one $a_{1g}$ and two $e_g'$
bands. The $a_{1g}$ and one of the two $e_g'$ bands cross the Fermi level, forming,
respectively, a large hexagonal hole pocket around the $\Gamma$ point, and six small,
elliptical hole pockets. The effect of hydration on the electronic structure has been shown
to be, for all practical purposes, related solely to lattice expansion \cite{MDJ04}, and we
will therefore use an unhydrated, but expanded compound for our calculations. The neglected
effects of Na or H$_2$O disorder would, if anything, further exaggerate two-dimensionality. 

	We have performed a tight-binding (TB) fit to our paramagnetic full-potential LAPW
\cite{Wien2k} band structure near E$_f$.  The bond length dependence of the TB parameters
was incorporated as described in Ref.\cite{DAP03}, and used to analyze Fermi surface 
dependence on the interlayer distance, $c$, with the apical O height set to its relaxed
value in the hydrated compound. The fits (details to be published elsewhere \cite{MDJ+})
have an rms error of only 3 mRy in the relevant energy range.  This level of accuracy and a
dense mesh of k-points (over 30,000 throughout the BZ)  were necessary for good
resolution of the very small FS hole pockets.  We use the TB Hamiltonian to calculate the
one-electron susceptibility $\chi_0({\bf q},\omega)= \chi_0'({\bf q},\omega)+ i\chi_0''({\bf
q},\omega)$, defined as $$ \chi_0({\bf q},\omega)=\sum_{\bf k}{[f(\epsilon_{\bf
k+q})-f(\epsilon_{\bf k})] /(\epsilon_{\bf k}-\epsilon_{\bf k+q}-\omega-i\delta)}, $$ where
$\epsilon_{\bf k}$ is the one-electron energy, $f$ is the Fermi function, and the matrix
elements are neglected \cite{entry-0}.  Fig. \ref{chi} exhibits the nesting structure 
in  
$\chi_0''({\bf q},\omega)/\omega$ at $\omega\rightarrow 0$,
important for superconductivity, and also in $\chi_0'(\mathbf{q},0)$.

\begin{figure} \begin{center} \includegraphics[width =
0.9\linewidth]{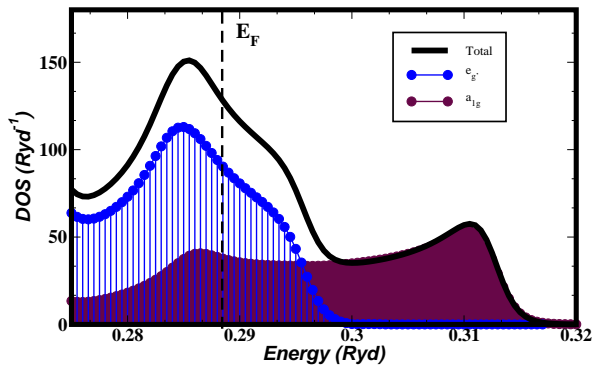}\end{center} \caption{(color online) The density of states of 
Na$_{0.3}$CoO$_2\cdot y
H_2O$ and of the two bands crossing the Fermi level, in the TB model \cite{MDJ+}.  The a$_{1g}$ band,
which carries 2/3 of all holes represents only about one third of N(E$_F$). }
 \label{DOS}
\end{figure}

\begin{figure}
 \begin{center} \includegraphics[width=0.9\linewidth]{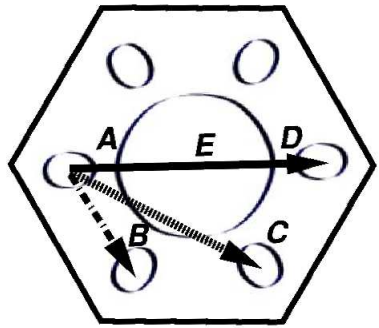}\end{center}
 \caption{The three main nesting types in 
 Na$_{0.3}$CoO$_2\cdot y H_2O$.  The perfect nesting at $\bf Q_1$ is shown as a solid 
line and the two
imperfect, $\bf Q_2$ and $\bf Q_3$, in narrow and wide dashed lines.} 
\label{nest} 
\end{figure}

\begin{figure} \includegraphics[width=0.9\linewidth]{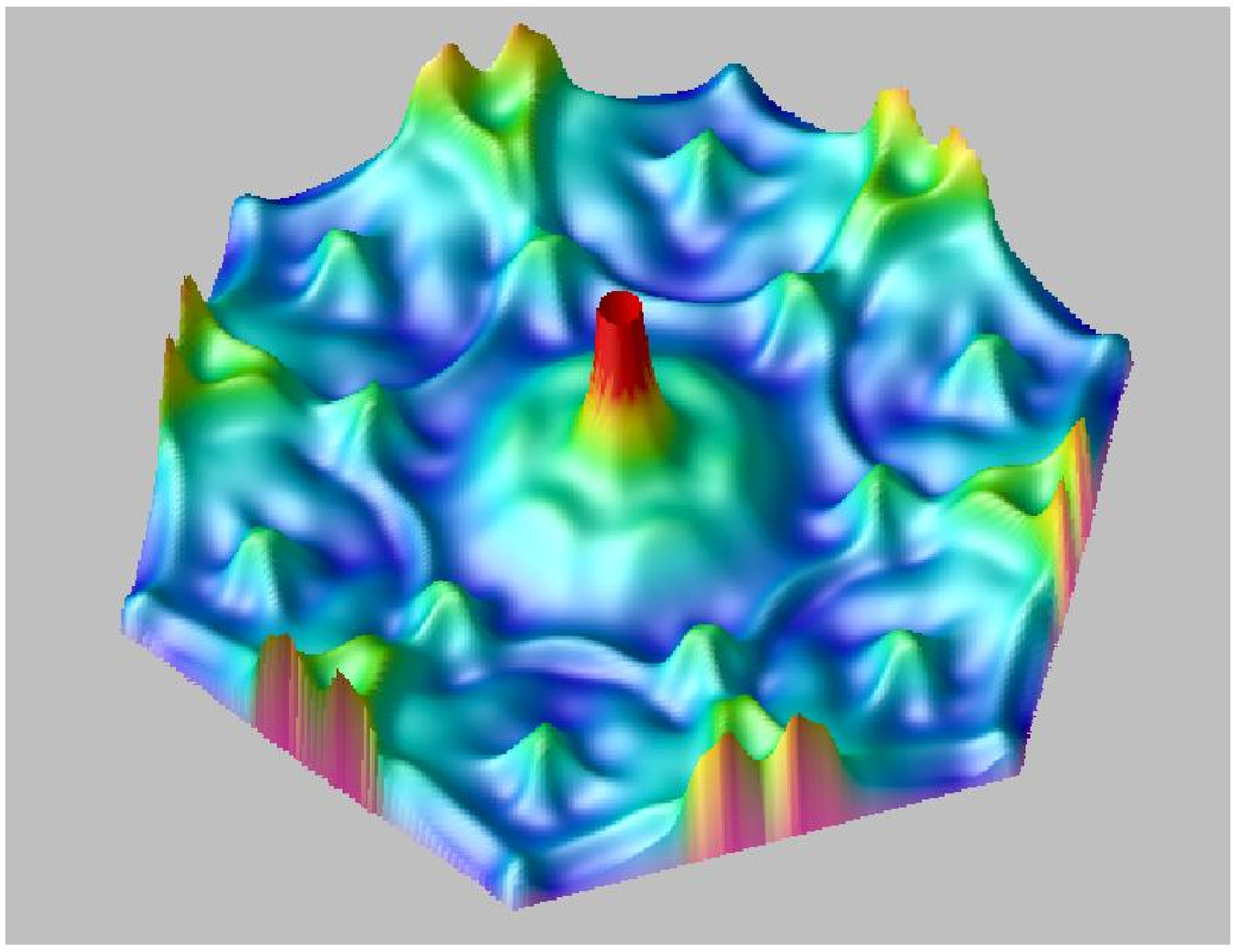}
\includegraphics[width=0.9\linewidth]{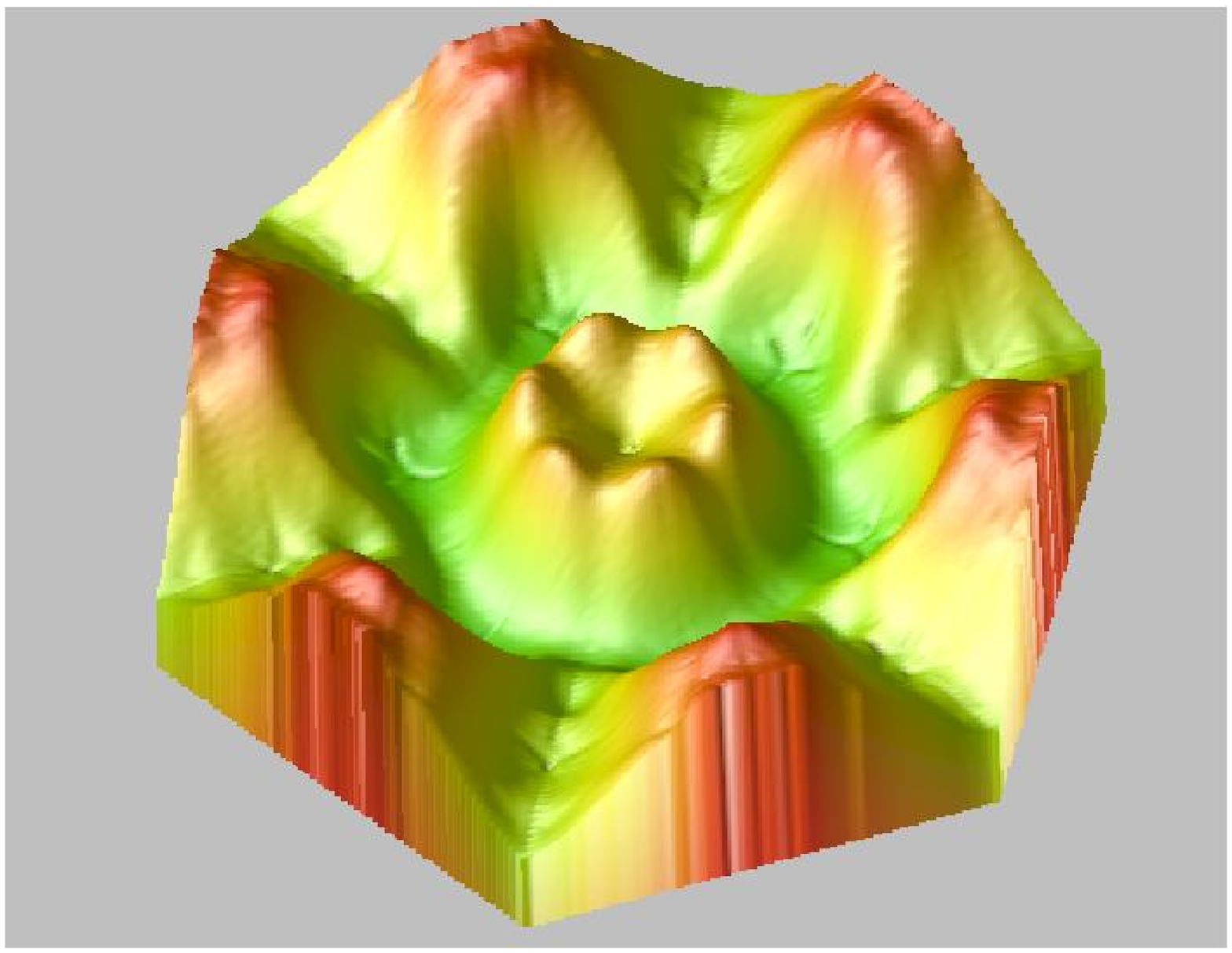}
\caption{ (color online) {\it top}: $\lim_{\omega\rightarrow 0}\chi''_0({\bf q,\omega})/\omega$ The 
double-humped peaks 
corresponding to $\mathbf{Q}_1$ nesting appear along the 
flat edges of the zone boundary. {\it bottom}:
$\chi'_0(\mathbf{q},0)$. A temperature broadening of .001 mRy was used.}
\label{chi} \end{figure}

With an increased $c$, as expected, the bands are completely two dimensional within the accuracy of our
calculations. The $e_{g'}$ holes get heavier, and comprise $\sim$ 70\% of the DOS at the Fermi energy (Fig.  
\ref{DOS}), though the total volume of these pockets is half the volume of the central $a_{1g}$ pocket. The
latter, which had formed a hexagonal prism with moderately flat faces in the parent compound, becomes nearly
circular and, hence, its contribution to the susceptibility is practically featureless. However, the six
elliptical pockets exhibit very good nesting and because of two dimensionality, their small size can only enhance
the susceptibility at the nesting vector, leaving the deviation from circular cylindrical shape as the only factor
determining nesting strength. Indeed, if they were exactly circular, all three nesting vectors $\bf Q_1$, $\bf
Q_2$, and $\bf Q_3$, shown in Fig. \ref{nest} by solid, narrow dashed and wide dashed lines, respectively, would
nest perfectly. No symmetry requirement forces these pockets to be circular, nor is there a restriction imposed on
their distance from the $\Gamma$ point. However, due to their small size they are nearly perfectly elliptical so
that, for example, pockets A and D in Fig. \ref{nest} nest nearly exactly, and, accidentally, the distance between
them is just slightly less than half of the reciprocal lattice vector $\bf G$. As a result, a strong double-humped
peak appears in the calculated $\chi_0''$ (Fig. \ref{chi}) at all half-integer reciprocal lattice vectors (the
distance between the humps measures the deviation of the nesting vector from ${\bf G}/2$).

Since $\mathbf{Q_1} \approx \mathbf{G}/2$, $\bf Q_2$ and $\bf Q_3$ are close to
$\mathbf{G}/4$, as seen in Fig.  \ref{chi}. The corresponding peaks are suppressed by
the ellipticity of the $e_{g'}$ pockets, which creates misorientation between the
pockets A and B or A and C. As long as the pockets are perfect ellipses, the peaks at
$\bf Q_2$ and $\bf Q_3$ have the same amplitude. In the real part of $\chi$ the peak at
$\bf Q_1$ is broadened but still prominent, but the peaks at $\bf Q_2$, $\bf Q_3$ are
smeared out.

{\it Ramifications for structure and superconductivity.} We have now established that there is prominent
nesting-related structure in both real and imaginary parts of the electronic susceptibility at all half-integer
reciprocal lattice coordinates, and weaker, but noticeable structure at all quarter-integer coordinates in the
imaginary part. Because of two-dimensionality \cite{2d}, in the charge channel this structure can lead to a
Peierls-type charge density wave instability, $i.e.,$ a structural transition. Indeed, various superstructures
have been reported, especially at $x=1/2,$ and are often ascribed to charge ordering of Na ions. Our results
suggest that the CoO$_2$ planes themselves have a tendency toward superstructure formation, even without Na
ordering. The observed superstructures, presumably, are affected by both factors. Finally, there are
indications \cite{QHMLF+04,BCS+04} that an antiferromagnetic ordering may set in parallel to a structural
instability, which in our picture would correspond to the condensation of a spin density wave at a nesting
vector.

At the compositions where the structure in $\chi' $ does \textit{not }lead to an
instability, one may expect soft modes associated with the corresponding wave vectors.
However, from the point of view of superconductivity, even more interesting are the
corresponding spin fluctuations, which take advantage not only of the structure in the
real part of $\chi_0 ,$ but also of the (much sharper)  structure of $\chi_0'' .$ Let
us first concentrate on the strongest peak in $\chi'' _{0},$ at $\mathbf{q=Q}_{1},$ and
consider possible signs of the order parameter on the corresponding pockets (A and D in
Fig.\ref{nest}). The relevant part of the linearized equation for the order parameter
can be written as:

\begin{equation}
\Delta(\mathbf{k}_A, i\omega_n) = T \sum_{\omega_n'} 
\frac{V(\mathbf{Q}_1, 
i\omega_n - i\omega_n')}{\xi^2_{\mathbf{k}_D} + |\omega_n'|^2} 
\Delta(\mathbf{k}_D, 
i\omega_n')
\label{gaps} 
\end{equation}

Here $\Delta $ stands for the order parameter which is scalar for singlet, and vector for
triplet pairing. The summation includes all Matsubara frequencies $\omega_n$, the
quasiparticle energies are given by $\xi_{\mathbf{k}}$, and $V(\mathbf{q})$ is the
pairing potential (positive for attraction).  For pairing induced by phonons,
$V(\mathbf{q})$ is always positive, while for spin
fluctuation exchange it is positive for triplet pairing
and negative for singlet. Since the pockets $A$ and $D$ are related
 by spatial inversion, for inversion symmetric
order parameters, $s,$ $d,$ $etc,$ a solution to Eq. \ref{gaps} exists 
only if $V(\mathbf{Q}_1)  >0,$ that is, in the
triplet channel, and for antisymmetric order parameters,
 $p$ $etc.,$ only in the singlet channel.  Note that such strict selection rules are related
to the small size of the $e_g'$ pockets.  Nesting in a large Fermi surface that can support
line nodes can, in principle, induce a singlet $d$-wave superconducting state, as discussed,
for instance, in Ref. \cite{YFHK03}.

\begin{figure} \includegraphics[width = 0.95\linewidth]{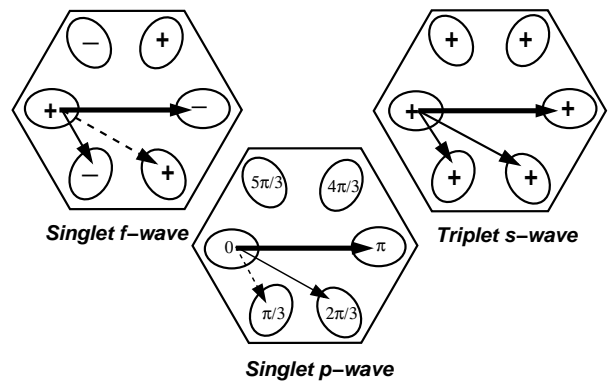} \caption{The order parameter phase on the 
relevant
sheets of the Fermi surface of various odd-gap superconducting states in NCO.
Solid black lines indicate pairing interactions and dashed lines pair-breaking. In the central panel, the phase
factors are e$^{i\delta}$ where $\delta$ is indicated in each pocket.}
\label{phase} 
\end{figure}

The Pauli principle forbids both singlet-$p$ ($Sp)$ and triplet-$s$ ($Ts)$ states, as
well as $Td$, unless the order parameter is odd with respect to Matsubara time, as
discussed first by Berezinski \cite{VLB74} for He$^{3}$, and in the context of solid
state by Balatsky and collaborators \cite{ABEA92}. In other words, spin fluctuations
with $\mathbf{q=Q}_{1}$ are pairing for the odd-gap $Sp$ state ($oSp)$ and for the
odd-gap $Ts$ state ($oTs).$ The relative stability of these must be decided by other
interactions. Staying within the spin-fluctuation scenario, we include now the two
other nesting vectors, $\mathbf{Q}_{2}$ and $\mathbf{Q}_{3}.$ Let us first consider the
$oS$ state. The order parameter is a scalar, and without loss of generality we assume
it to be constant within each of the pockets, with a phase changing from pocket to
pocket as in Fig \ref{phase}. The lowest angular momentum solution allowed for a real
order parameter corresponds to an $f$-wave, changing among the pockets as $\Delta
_{A}:\Delta _{B}:\Delta _{C}:\Delta _{D}=1:-1:1:-1, $ which implies several node
lines in the induced gap on the pocket E. A complex $p$-wave order parameter is also
allowed, so that $\Delta _{A}:\Delta _{B}:\Delta _{C}:\Delta _{D}=1:e^{i\pi
/3}:e^{2i\pi /3}:-1.$ In this case, the induced order parameter on the pocket E varies
with angle as $e^{i\phi},$ and is nodeless, and thus energetically more favorable. Note
that as long as the spin-fluctuations with vectors $\mathbf{Q}_{2}$ and
$\mathbf{Q}_{3}$ couple to electrons with equal strength, they do not contribute to
pairing at all in either case, canceling each other completely (because of the phase
factor).

On the other hand, the $oTs$ state can actually benefit from the two \textquotedblleft minor\textquotedblright\
nestings. In this case, there are no symmetry-implied nodes, and the simplest solution has constant gaps over
both types of Fermi surfaces. The order parameters are now spinors, which can be translated in a standard way
into real-space vectors. Several solutions are possible with pair spins either in or perpendicular to the
hexagonal plane, for instance, $\mathbf{d}=const\cdot \mathbf{\hat{z}},$ where $\mathbf{\hat{z}}$ is the unit
vector in $c$ direction, or \textbf{d}$=const\cdot (\mathbf{\hat{x}+}i\mathbf{\hat{y}).}$ Importantly, for any
of the $oTs$ states, all spin fluctuations are pairing, including those with $\mathbf{q=Q}_{2}$ and
$\mathbf{q=Q}_{3}.$ Furthermore, an $oTs$ state seems compatible with the limited experimental information
available on the pairing symmetry. First, such a state would not exhibit exponential behavior in DOS-probing
experiments (specific heat, NMR/NQR, penetration depth) \cite{TFGZ+,KIYI+}. Second, a coherence peak in the NMR
relaxation rate $T_{c},$ if it exists, should be suppressed compared to a conventional $eSs$ state. In most
experiments such a peak is not observed, but a rather weak coherence peak was seen in Ref.  \cite{YKMY,TWCM+}.
Third, specifically the $\mathbf{d}\parallel \mathbf{\hat{z}}$ triplet $s$-state, similar to the chiral
$p$-state in Sr$_{2}$RuO$_{4}, $ implies no change in the in-plane susceptibility, as measured by the Knight
shift. \cite{WHKO+03,AKAK+03, TWCM+}.  Finally, non-magnetic impurities are not pair-breaking for an $oTs$
state, in agreement with observation \cite{MYHW+03}.

{\it Conclusion}. We calculated the bare susceptibility of Na$_{1/3}$CoO$_2$ in the
expanded structure, corresponding to Na$_{1/3}$CoO$_2\cdot y $H$_2$O. We found that
the main contribution to the density of states comes not from the large Fermi
surface pocket at the zone center, but from small surrounding pockets. These pockets
exhibit strong nesting features, accidentally commensurate with the lattice, and
suggest proximity to both charge and spin density wave instabilities.
The structure of the spin fluctuations is such that it favors an odd-gap triplet
$s$-wave superconductivity, which seems compatible with existing
experiments. The results reported in this Letter are calculated for $x$=0.3,
but we have verified that the described nesting features remain
up to at least $x$=0.5.

{\it Acknowledgments}.	We acknowledge valuable discussions with A. Balatsky,
 D. Mandrus, S. Nagler, and B. Sales. We are particularly grateful to D. Agterberg
 for pointing out to us the possibility of odd-gap superconductivity in connection with the
 nesting in this compound.


\begin{thebibliography}{29}
\bibitem[{ent()}]{entry-0}
\bibinfo{note}{This approximation is good as long as the character of the wave
  functions at {\bf k} and {\bf k+q} is the same. This is true for $a_{1g}$
  transitions, as well as for the $\bf Q_1$ nesting. Because of the double
  degeneracy of the $e_{g'}$ representation, the matrix elements will be
  reduced for $\bf Q_2$ and $\bf Q_3$ by a factor of up to two, and strongly
  suppressed for the weakly-structured $a_{1g}-e_{g'}$ component. For the
  qualitative discussion here, however, the constant matrix elements
  approximation is sufficient.}
\bibitem[{2d()}]{2d}
\bibinfo{note}{Note that Fig. 3 applies to hydrated NCO with x=0.3. Unhydrated
  NCO is less two-dimensional, and even less as x increases. However, we expect
  the unhydrated compound still demonstrates similar nesting properties albeit
  less pronounced.}
\end{thebibliography}
\end{document}